# Comment on "The rise of semiconductor spintronics"
# (ArXiv:0904.3034)

*Vladimir L. Korenev,*

*Division of Solid State Physics,
Ioffe Physical Technical Institute, Politekhnicheskaya 26,
St. Petersburg, 194021 Russia; tel.007 (812) 2927396,
Korenev@orient.ioffe.ru*

The article "The rise of semiconductor spintronics" (*Nature Physics* **4** S20 2008, http://www.nature.com/milestones/spin, milestone 23) essentially says that Awschalom and co-workers have done all the major experiments that have born semiconductor spintronics. This is far from being true, although certainly their works are very interesting and important. There was a long history of research on spin properties long before the word "spintronics" was invented. Many important works are now forgotten, so I would like to argue that most of the key experiments were done and published earlier than the papers cited in the article.

Four prominent experimental results have been chosen:

**(1) Long spin coherence (spin memory) times in semiconductors**

**(2) Spin transport over macroscopic distances**

**(3) Injection of spin from magnetic substances into non-magnetic semiconductors**

**(4) Electrical generation and manipulation of spin in non-magnetic semiconductors**

**Item (1)** dates back to Ref. [1] that reported 30 ns spin memory time in n-GaAs 20 years earlier than PRL **80** 4313 (1998) cited by the article. The first publication [2] of long term spin memory (42 ns) in a refereed journal did not pretend to discover the long spin relaxation time. Its main result was the first report of **Item (2)** – macroscopically large spin diffusion length in GaAs, definitely obtained before Nature **397** 139 (1999). **Item (3)** was first demonstrated in [3]. **Item (4):** Current induced spin polarization in a semiconductor was first demonstrated in [4]. Electrical manipulation of the non-equilibrium spin in semiconductors was realized for the first time in [5]: the electron drift in an externally applied electric field induces the rotation of optically injected spin by means of spin-orbit interaction. The Spin Hall effect (SHE) was observed experimentally for the first time by Awschalom's group [6] and practically simultaneously in Ref. [7]. However, one should recognize the earlier experiments [8, 9] on the Anomalous Hall Effect (AHE), which is reciprocal to SHE.

Below I consider each of the items in more detail.

# LONG SPIN COHERENCE (SPIN MEMORY) TIMES IN SEMICONDUCTORS

Weisbuch [1] measured spin lifetime in n-type GaAs with the use of the Hanle effect. In this case circularly polarized light creates electrons with preferential spin orientation. If the electron spin relaxation time $\tau_S$ is not too short in comparison with the lifetime $\tau_J$, a spin polarization builds up.

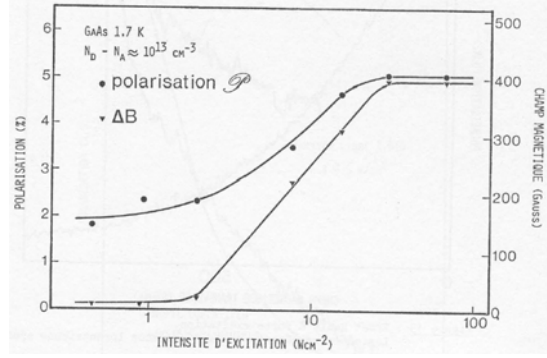

**Figure 1. Power dependence of the halfwidth Hanle curve (triangles, right scale) and zero field electron spin polarization (circles, left scale). Adapted from the Ref.1.**

If an external magnetic field $\vec{B}$ is applied perpendicular to the pump beam, the average electron spin precesses around the field with the Larmor frequency $\mu_B g_e B/\hbar$ ($g_e$ is the electron $g$ factor, and $\mu_B$ is the Bohr magneton). In steady state, the electron spin projection $S_z$ on the initial direction decreases with increasing field, which brings about PL depolarization (the Hanle effect). The $S_z(B)$ dependence is described by the relation

$$S_z(B) = 0.25 \frac{\tau_S}{\tau_S + \tau_J} \frac{1}{1 + (B/\Delta B)^2}, \qquad (1)$$

where the reciprocal spin lifetime

$$T_s^{-1} = \tau_s^{-1} + \tau_J^{-1} \qquad (2)$$

can be determined by measuring the half-width of the magnetic depolarization curve $\Delta B = \hbar/\mu_B g_e T_s$. Eqs (1) and (2) enable one to determine times separately. In n-type semiconductors the lifetime $\tau_J$ of electons (majority charge carriers) is much longer than $\tau_S$ in a weak-pumping limit. In this case the spin lifetime $T_S$ is equal to the electron spin-relaxation time $\tau_S$. Measurements of zero-field polarization (circles on Fig.1) and the Hanle curve halfwidth (tiangles on Fig.1) enabled Weisbuch to obtain $\tau_s \approx 30\,ns$ in the weak-pumping limit.

# SPIN TRANSPORT OVER MACROSCOPIC DISTANCES

By macroscopic distance I mean the distance larger than the typical size of microelectronic element, i.e. larger than 1 micron. The long-term spin memory of n-type GaAs (for low doping $\sim 2 \cdot 10^{15}\, cm^{-3}$) enabled Dzhioev et al [2] to measure for the first time the macroscopic spin diffusion length $L_S$. If the photo-holes are distributed deep into semiconductor more or less uniform by the photon recycling process the diffusion of electron spin takes place with the high electron diffusion coefficient $D_e = 24\, cm^2/s$, which together with the long $\tau_S = 42\, ns$ gives the long $L_s = \sqrt{D_e \tau_s} = 10\,\mu$. The spectral dependence of PL polarization $\rho_c$ reflects the electron spin diffusion into semiconductor. Indeed, a photon created in recombination of an electron and a hole at a distance $z$ from the semiconductor surface escapes from the sample with the probability $\exp[-\alpha(\lambda)z]$. The absorption coefficient $\alpha(\lambda)$ depends strongly on the wavelength $\lambda$ of emission. The absorption coefficient at the high-energy wing of the PL line is high, and one detects the polarized light which has been emitted near the sample surface. In contrast, at the low-energy wing of PL (where $\alpha(\lambda)$ is low), recombination of all electrons can be seen. This results in the reduction of $\rho_c$ in the low-energy part of the line, since the spin density in the bulk of the sample is lower than that at the surface. The zero-field spectral dependence of the PL polarization is given by formula

$$\rho_c(\lambda) = \alpha \int_0^\infty s(z) e^{-\alpha z} dz = s(z=0) \frac{\alpha(\lambda) L_s}{1 + \alpha(\lambda) L_s} \qquad (3)$$

where the mean spin $s(z=0)$ at the surface and the spin diffusion length $L_s$ are the fitting parameters. Comparing the $\rho_c(\lambda)$-dependence measured by optical orientation technique with well-known $\alpha(\lambda)$-dependence for GaAs of similar doping (Fig.2), one finds $L_s = 10\,\mu$.

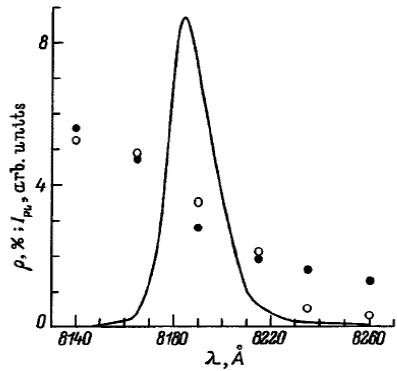

**Figure 2.** Spectra dependence of PL intensity (solid line), circular polarization $\rho_c$ (filled circles) and the plot of Eq.(3) (open circles). After Ref.2.

# INJECTION OF SPIN FROM MAGNETIC SUBSTANCES INTO THE NON-MAGNETIC SEMICONDUCTOR

Alvarado and Renaud [3] were the first to observe spin injection. Electric current from the ferromagnetic (nickel) tip of scanning tunneling microscope (STM) into p-GaAs is acompanied by the spin current. The electrically injected spin-polarized electrons recombine with holes in p-GaAs bringing about electroluminescence (Fig.3). According to selection rules, the electroluminescence becomes circularly polarized, with the degree $\rho_c$ being proportional to the spin injection efficiency $P_M$. Alvarado and Renaud found the efficiency $P_M$ as high as 48%.

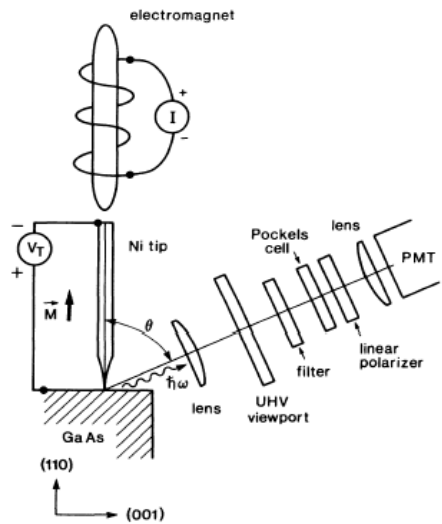

**Figure 3. Scheme of experiment on spin injection from STM Ni tip into GaAs. Adapted from the Ref.3.**

# ELECTRICAL GENERATION AND MANIPULATION OF SPIN IN NON-MAGNETIC SEMICONDUCTORS

(i) Current induced spin polarization in semiconductor was first demonstrated by Vorobjev *et al* [4]. They detected the spin-polarization-induced Faraday rotation of the plane of polarization of light propagated in a tellurium crystal. The spin polarization of valence band electrons (holes) was created under electric current flow along the major axis $C_3$ (z-axis). This effect is due to the spin-orbit interaction, which is responsible for the unique valence band structure in Te (Fig.4). The upper valence band is non-degenerate: at the states with momentum $k_z > 0$ the projection of angular momentum $m_z = +3/2$ contributes mainly to the wavefunction of electron, whereas $m_z = -3/2$ dominates for the $k_z < 0$ states. However, there is no spin polarization in equilibrium. When an electric current flows the Fermi level is tilted as shown in Fig.4. The redistribution of electrons in $\vec{k}$-space leads immediately to the spin polarization of the valence band electrons. The electrically induced redistribution of electrons over the valence band states changes the absorption coefficients of the right and the left circularly polarized light, thus leading to the rotation of the light polarization plane.

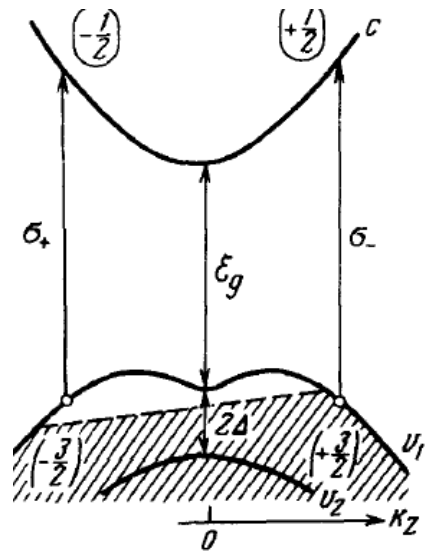

**Figure 4. Band diagramm for Te crystal. Dashed line shows the Fermi level in the valence band. Adapted from the Ref.4.**

(ii) The electrical manipulation of electron spin direction in semiconductors was realized for the first time by Kalevich and Korenev [5]: the electron drift induces the electron spin rotation, which scales with an external electric field. Indeed, the zero-field spin splitting of the conduction band can be considered as an interaction of electron spin with an effective magnetic field $\vec{H}_{eff}(\vec{k})$ whose value and direction are determined by those of the electron momentum $\vec{k}$. In low symmetry semiconductors (quantum wells, strained bulk GaAs etc) the field is linear in $\vec{k}$. As a result of momentum scattering the average effective field $\langle \vec{H}_{eff} \rangle$ is absent in equilibrium. The electric current induces an electron drift with the velocity $\vec{v}_d$, resulting in a non-zero average field $\langle \vec{H}_{eff} \rangle \propto \vec{v}_d$. The $\langle \vec{H}_{eff} \rangle$-field produces spin precession, which is similar to the Larmor precession of electrons in external magnetic field $\vec{H}$. As a result, the Hanle effect takes place in a total field $\vec{H} + \langle \vec{H}_{eff} \rangle$. This leads to the shift of the Hanle curve by the $\langle H_{eff} \rangle$ value if $\langle \vec{H}_{eff} \rangle \| \vec{H}$. The measured Hanle curve asymmetry is shown in Fig.5 (the low-field cusp may come from localized states for which $\langle H_{eff} \rangle = 0$).

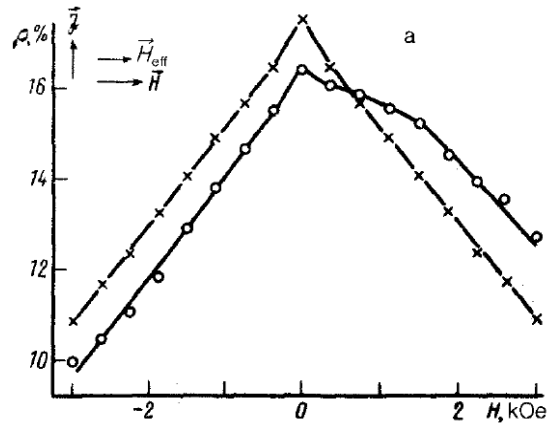

**Figure 5. The Hanle effect in the absence (crosses) and in the presence (circles) of electric current. Adapted from the Ref.5.**

(iii) The Spin Hall effect (SHE) and the Anomalous Hall Effect (AHE) are the two sides of the same medal (Fig.6). In both cases electrons with up and down spins deflect perpendicular to their drift direction due to spin-orbit interaction. AHE (Fig.6a) consists in the appearance of Hall current $\vec{j}_{AHE} \propto \vec{E} \times \vec{S}_{bulk}$ in the presence of flow of spin-polarized electrons with the bulk spin density $\vec{S}_{bulk}$. The AHE effect was observed in a semiconductor by Chazalviel and Solomon [8].

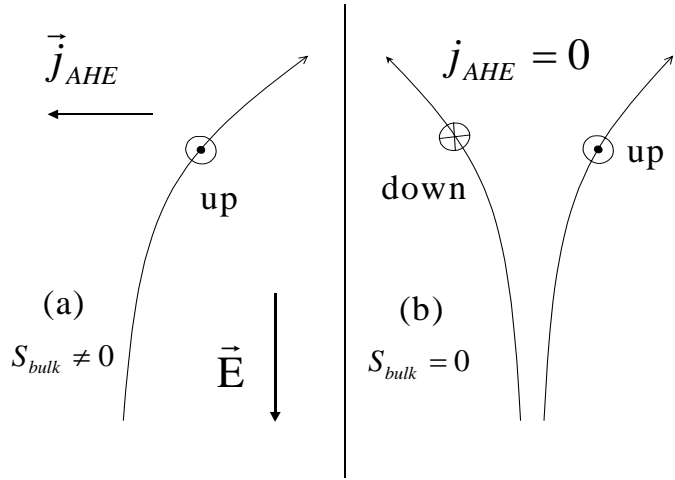

**Figure 6. Illustartion of (a) the Anomalous Hall Effect and (b) the Spin Hall effect**

The Hall current is absent if $S_{bulk} = 0$ (Fig.6b). There is however non-zero *spin* Hall current originating from the correlation between spin and momentum. It leads to the spin accumulation at the left and the right boundaries of the sample – the Spin Hall effect. Awschalom's group [6] and Wunderlich et al [7] observed the SHE for the first time. This is important observation.

It is interesting to note that not only drift but also the directed diffusive motion results in similar effects. The diffusive version of the AHE was observed by Bakun et al [9]. To the best of my knowledge, nobody reported the diffusive version of the SHE.